\documentclass[11pt]{article}
\usepackage{graphicx}
\usepackage[margin=1.25in]{geometry}
\usepackage[usenames,dvipsnames]{color}
\usepackage{url}
\usepackage[colorlinks = true,
            linkcolor = blue,
            urlcolor  = blue,
            citecolor = blue,
            anchorcolor = blue]{hyperref}
\usepackage{tcolorbox}

\newcommand\pubdate{\today}

\def\Title#1{\begin{center} {\LARGE #1 } \end{center}}
\def\Author#1{\begin{center}{ \sc #1} \end{center}}
\def\Address#1{\begin{center}{ \it #1} \end{center}}

\newcommand\pubblock{\rightline{\begin{tabular}{l}
         \pubdate \end{tabular}}}
\newenvironment{Abstract}{\begin{quotation} \begin{center}
                       ABSTRACT
     \end{center}\bigskip  }{\end{quotation}}

\def\beq{\begin{equation}}
\def\eeq#1{\label{#1}\end{equation}}
\def\eeqn{\end{equation}}

\newenvironment{Eqnarray}%
   {\arraycolsep 0.14em\begin{eqnarray}}{\end{eqnarray}}
\def\beqa{\begin{Eqnarray}}
\def\eeqa#1{\label{#1}\end{Eqnarray}}
\def\eeqan{\end{Eqnarray}}

\let\bar=\overbar

\def\lsim{\mathrel{\raise.3ex\hbox{$<$\kern-.75em\lower1ex\hbox{$\sim$}}}}
\def\gsim{\mathrel{\raise.3ex\hbox{$>$\kern-.75em\lower1ex\hbox{$\sim$}}}}

\def\del{\partial}
\def\Dslash{\not{\hbox{\kern-4pt $D$}}}
\def\dslash{\not{\hbox{\kern-2pt $\del$}}}
\def\pslash{\not{\hbox{\kern-2pt $p$}}}
\def\ETmiss{\not{\hbox{\kern-4pt $E$}}_T}

\def\Dlr{\mathrel{\raise1.5ex\hbox{$\leftrightarrow$\kern-1em\lower1.5ex\hbox{$D$}}}}

\def\MSB{{\bar{M \kern -2pt S}}}
\def\msb{{\bar{\scriptsize M \kern -1pt S}}}

\def\drb{{\bar{\scriptsize D \kern -1pt R}}}

\newcommand\snowmass{\begin{center}\rule[-0.2in]{\hsize}{0.01in}\\\rule{\hsize}{0.01in}\\
\vskip 0.1in Submitted to the  Proceedings of the US Community Study\\ 
on the Future of Particle Physics (Snowmass 2021)\\ 
\rule{\hsize}{0.01in}\\\rule[+0.2in]{\hsize}{0.01in} \end{center}}

\begin{document}

\pubblock

\Title{When, Where, and How to Open Data \\\vspace{3mm} \textit{A Personal Perspective}}

\bigskip 

\Author{Benjamin Nachman}

\medskip

\Address{Physics Division, Lawrence Berkeley National Laboratory, Berkeley, CA 94720, USA \\ Berkeley Institute for Data Science, University of California, Berkeley, CA 94720, USA}

\medskip

 \begin{Abstract}
\noindent This is a personal perspective on data sharing in the context of public data releases suitable for generic analysis.  These open data can be a powerful tool for expanding the science of high energy physics, but care must be taken in when, where, and how they are utilized.  I argue that data preservation even within collaborations needs additional support in order to maximize our science potential.  Additionally, it should also be easier for non-collaboration members to engage with collaborations.  Finally, I advocate that we recognize a new type of high energy physicist: the `data physicist', who would be optimally suited to analyze open data as well as develop and deploy new advanced data science tools so that we can use our precious data to their fullest potential.\\

\noindent This document has been coordinated with a white paper on open data commissioned by the American Physical Society's (APS) Division of Particles and Field (DPS) Community Planning Exercise (`Snowmass') Theory Frontier~\cite{other} and relevant also for the Computational Frontier.
\end{Abstract}

\snowmass

\def\thefootnote{\fnsymbol{footnote}}
\setcounter{footnote}{0}

\clearpage
\section{Introduction and Overview}

There is increasing interest across many disciplines in ``open science''. This can refer to sharing analysis code, data, experimental machinery, or protocols with people other than the scientists who did the original work. It can be motivated by a desire to archive the data for later re-analysis, to drive a faster pace of discovery by involving more scientists, or to ensure reproducibility and transparency in science. How to produce policies from these arguments is an ongoing challenge -- one in which modern high energy physics (HEP) experiments can be instructive\footnote{See Ref.~\cite{other} for a brief historical context.}, as they offer extreme examples of dataset and collaboration size. 

Typical HEP experimental papers have hundreds to thousands of authors. We have this tradition because data ``collection and curation'' includes detector design and operation, simulation development, pattern recognition, and calibration, and is a significant part of our collective work and is highly integrated with data analysis.  While the growing divide between ``theorists'' and ``experimentalists'' is worrisome for the future of HEP, it currently has important implications for funding and career advancement.  The HEP model is different from nearby fields with a tradition of open data.  For example, the researchers who build and operate detectors in astronomy are often not the people who analyze the data.

It is also worth noting that our field has a long history of data sharing, where anyone in the collaboration (a considerable fraction of all HEP physicists) has access to the raw data. We also have advanced procedures for collaborating with external scientists and sharing data for targeted usage via tools such as \texttt{HepData}~\cite{Maguire:2017ypu}.  While this setup has been very successful, it is being challenged by the recent public release of data suitable for generic data analysis by the CMS collaboration~\cite{cmsopen} (to be followed by the other LHC experiments~\cite{CERN-OPEN-2020-013}).  Henceforth, ``open data'' will refer to data suitable for generic data analysis and not a derived data product like digitized figures or tables.  I will also mostly be referring to open data in the context of running experiments, although I will explore the connection with legacy data analysis as well.

The motivation for open data seems to be two-fold: ensuring data preservation and increasing the chances for discovery. Data preservation is of paramount importance and is critically understaffed for ongoing and finished experiments.  Releasing and stress-testing data during the lifetime of an experiment is certainly one way to achieve data preservation.  However, this is not the only way. Data preservation can also be accomplished within the collaboration by ensuring that all older data can be analyzed.  This can be stress-tested automatically with continuous integration workflows or manually by performing physics analysis on older data, as was done in the case of the recent $W$ boson mass measurement from the ATLAS collaboration using early Run 1 data~\cite{ATLAS:2017rzl}.  It is not a given that older data are readily accessible internally; in fact, I would be surprised if any new graduate student on ATLAS would be able to readily interrogate the data from Run 1 as data formats and software tools have changed significantly.  This is a serious problem that should be addressed independent of data releases.  Some studies with the CMS open data~\cite{Komiske:2019jim,Lassila-Perini:2021xzn,other} have discussed the challenges with using these data.  These stories sounded painfully familiar of the typical graduate student struggle.  While most students do not get to write a paper about this struggle (it usually does not even make their thesis), their challenge is no less important and indicative of a serious challenge for data access and preservation internal to experimental collaborations.  This does not have to be a necessary part of the graduate student experience.  

In terms of discovery, having more ``eyes on the data'' is not obviously a problem for our field - the collaborations at the Large Hadron Collider (LHC) have many thousands of scientists and there are well-tested mechanisms for external scientists (including theorists) to join for generic (by becoming an author) or limited-scope projects.  Proposals for new ideas can be readily carried out with simulations (even ``open simulations'' if the situation calls for it).

\begin{figure}[h!]
    \centering
    \includegraphics[width=0.8\textwidth]{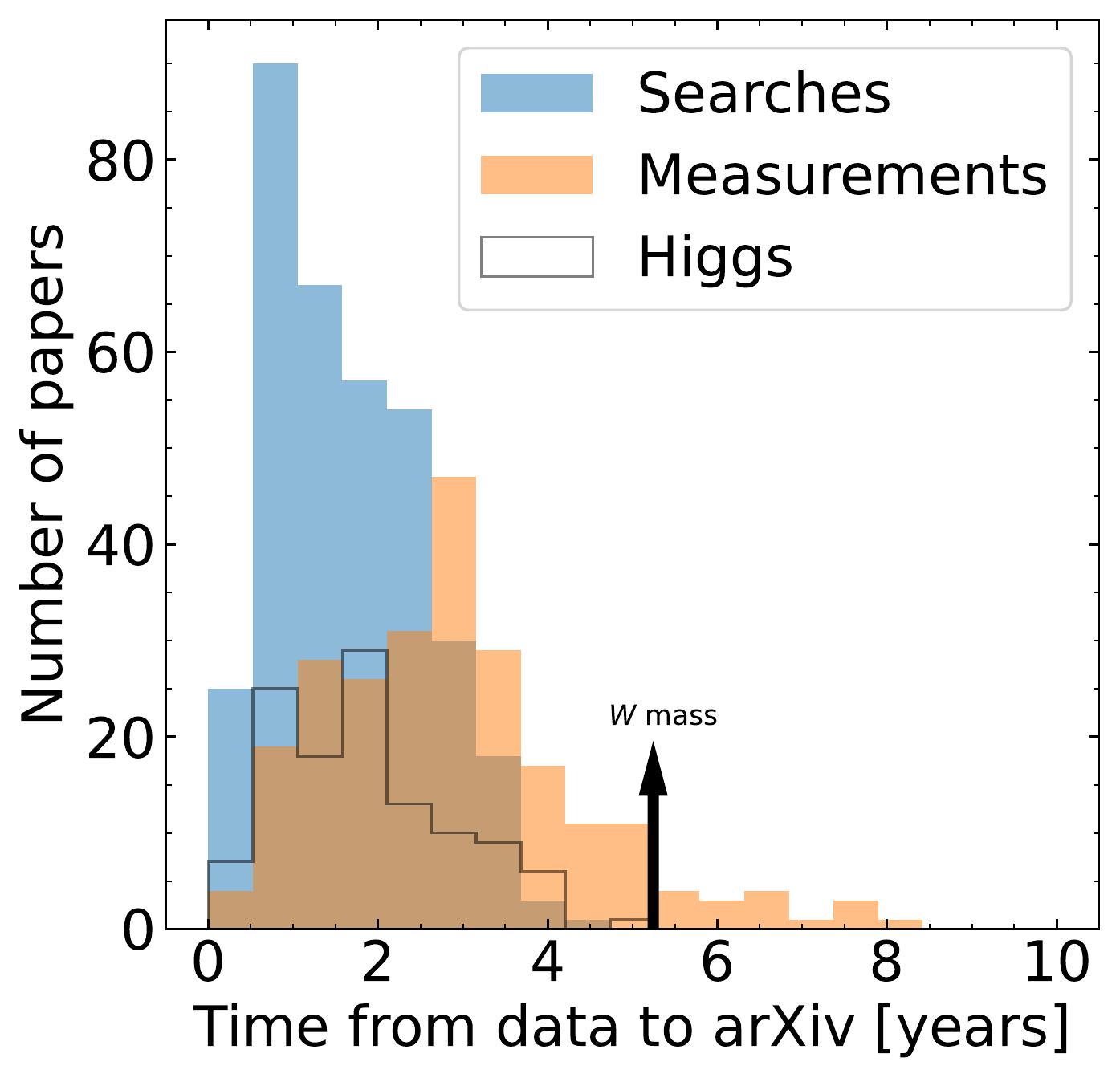}
    \caption{A histogram showing the time between the end of data taking and the posting of a paper to arXiv for physics papers from the ATLAS Collaboration, scraped from \url{https://twiki.cern.ch/twiki/bin/view/AtlasPublic}.  `Searches' include publications from the SUSY, Exotics, and HDBS groups while `Measurements' includes publications from the Standard Model, Top Quark, and B Physics groups.  Datasets less than 1 fb${}^{-1}$ are excluded.  There are 703 papers in the above histogram.}
    \label{fig:delays}
\end{figure}

While it may not be necessary for data preservation and for broadening HEP science, open data could advance both of these areas.  Along with these potential benefits, there may also be potential costs.  If the data are made available to those who are not involved in the ``collection and curation'', what is the incentive to take part in these important activities?  If the incentive to collect complex data is removed, the complex data may never be collected. Furthermore, those not involved in the ``collection and curation'' will likely not be able to perform accurate and precise experimental measurements because they are not familiar with the intricate details of the data.  Time delays are often a mechanism proposed to protect the incentive to work on data collection and curation.  However, any time delay other than the lifetime of an experiment is a statement about what sort of science is most ``interesting''.  Multi year delays would likely protect high profile, statistics-limited analyses.  Systematic uncertainty-limited analyses would not be protected (e.g. precision Standard Model measurements).  See Fig.~\ref{fig:delays} for the actual time delays between data collection and paper completion - there is a clear shift between searches and measurements with a heavy tail to the right for the latter.  It has also not been long enough since the end of Run 2 to know what the true tail will look like in the future.

It is often stated that open data would level the playing field and make HEP science more equitable.  However, open data may also have the opposite effect.  Large collaborations have membership from a diverse set of institutes who do not all have the same local resources.  Analyzing HEP data requires significant resources (computational and otherwise).  Working within a HEP collaboration ensures that all members have a chance to analyze these data.  In contrast, open data is only analyzable by wealthy groups (where ‘wealth’ has multiple meanings).  There is also the unfortunate reality that researcher diversity decreases with seniority (see e.g. Ref.~\cite{ATL-GEN-PUB-2016-001}) while the bulk of the collection and curation work within collaborations is performed by relatively early career researchers.  

Lastly, open data may reduce the integrity of HEP science if analysis scope and reviewer expectations are not clear and calibrated.  The release of CMS open data has already created at least one incident where a paper from a collaboration was initially rejected from a journal because the reviewers believed the analysis had already been done by theorists working on CMS open data~\cite{1711.08341}.  Open data makes it ``easy'' to compare predictions to data, but if those data are not carefully corrected for detector effects or if uncertainties are not quantified, then the comparison may not be meaningful.  A study comparing uncorrected data with predictions is not the same as a proper “measurement”.  The word ``measurement'' is in quotes because it carries a certain gravitas with ``experimentalists'' that it often does not with ``theorists''.   This could be addressed with a more integrated training of ``experimentalists'' and “theorists” in the future (see below).  It is also worth noting that many applications of open data could simply use simulation (or even open simulation) - the data themselves do not add to the scientific aspect of the research.  Once again, this is often the case for papers that propose new methods.  There is also a difference between `performance studies\footnote{It is unfortunate that these studies are often only findable on collaboration wiki pages or at best, the CERN Document Server. This is changing with many studies findable on \textsc{Inspire}, but it would increase idea sharing across the community to also post them to \textsc{arXiv}.}' (e.g. Refs.~\cite{ATLAS-CONF02014-048,ATLAS-CONF-2013-086}) and `measurements' (e.g. Refs.~\cite{1506.05629,1509.05190}).

\clearpage

\section{Experience with Diverse Open Data Models}

This section briefly illustrates my personal experiences with different models of data sharing.

\paragraph{ATLAS.}  ATLAS does not yet make data available for generic data analysis, although a number of datasets are available for outreach~\cite{ATL-OREACH-PUB-2020-001} and a growing number of open simulations have been published (see e.g. Ref.~\cite{fastcalogan}).  The goal of this section is to describe the experience (from the ``experimentalist'' perspective) of ``theorists'' working directly with ATLAS.  This is done through short term associate (STA) positions that are approved by the collaboration management~\cite{sta}.  I have worked with STAs on a number of occasions~\cite{1506.05629,2005.02983}.
The threshold to set up one of these positions is quite low and readily allows for the ``theorist'' to interact with internal data/simulation. This was highly effective. It would be useful to hear the perspective from the ``theorists'' to see if their experience was as positive.

\paragraph{CMS.}  I do not have any direct experience analyzing CMS Open Data, but I would completely believe the stories in Refs.~\cite{Komiske:2019jim,Lassila-Perini:2021xzn,other} about how difficult it is to analyze them from the public releases (despite the impressive effort by CMS to make it as easy as possible).  On the other hand, the simplified data formats made available through the \texttt{EnergyFlow} package (\url{https://energyflow.network}) make it very simple to access these data after some processing.  I have used the corresponding simulations in a number of studies~\cite{2107.08979,2112.05722,2205.03413}, where having full detector simulations was actually necessary to make a point.  The \texttt{EnergyFlow} package is a great service to the community and certainly lowers the barrier to entry for Open Data/Open Simulation studies.

\begin{figure}
    \centering
    \includegraphics[width=0.49\textwidth]{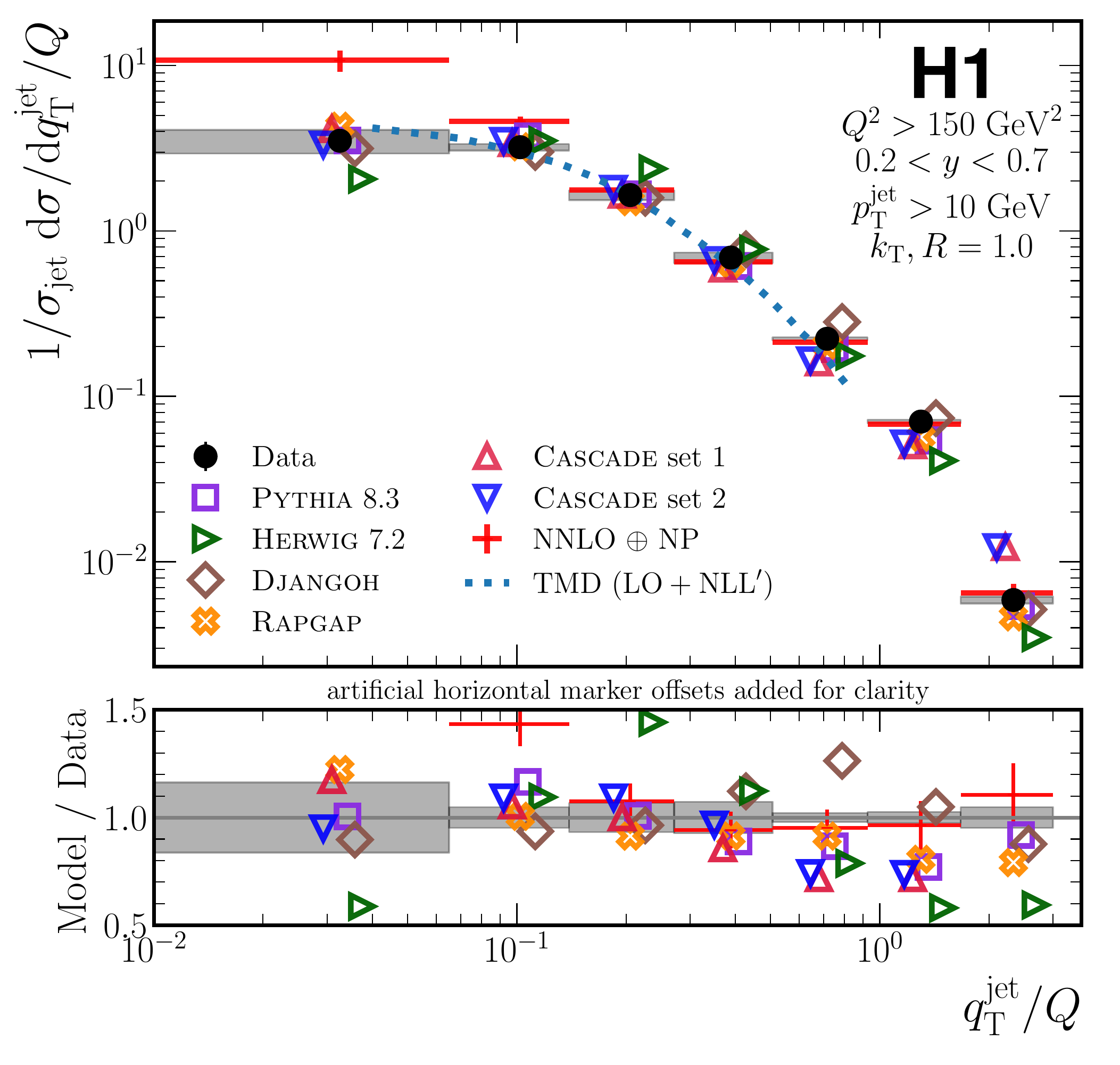}
    \includegraphics[width=0.49\textwidth]{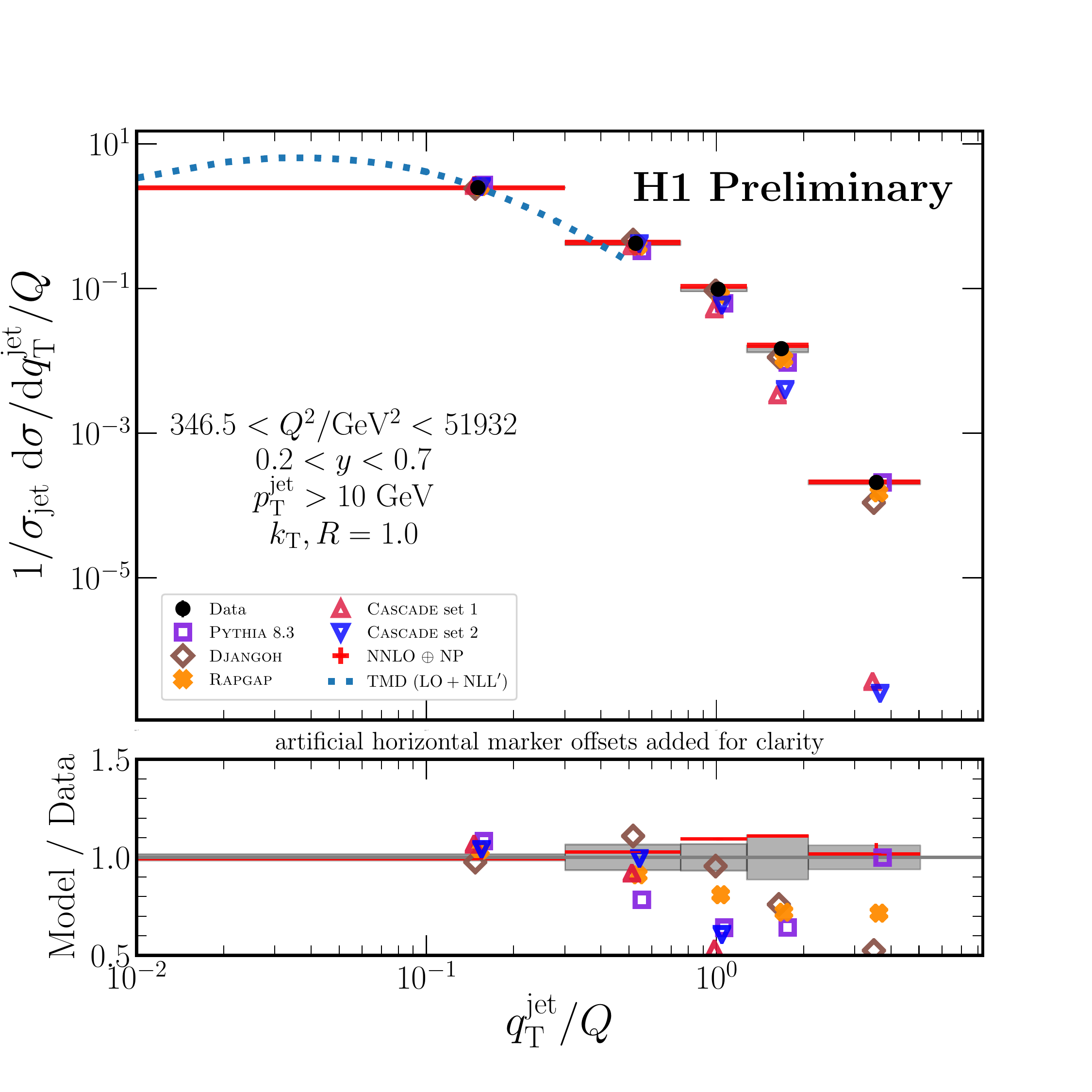}\includegraphics[width=0.49\textwidth]{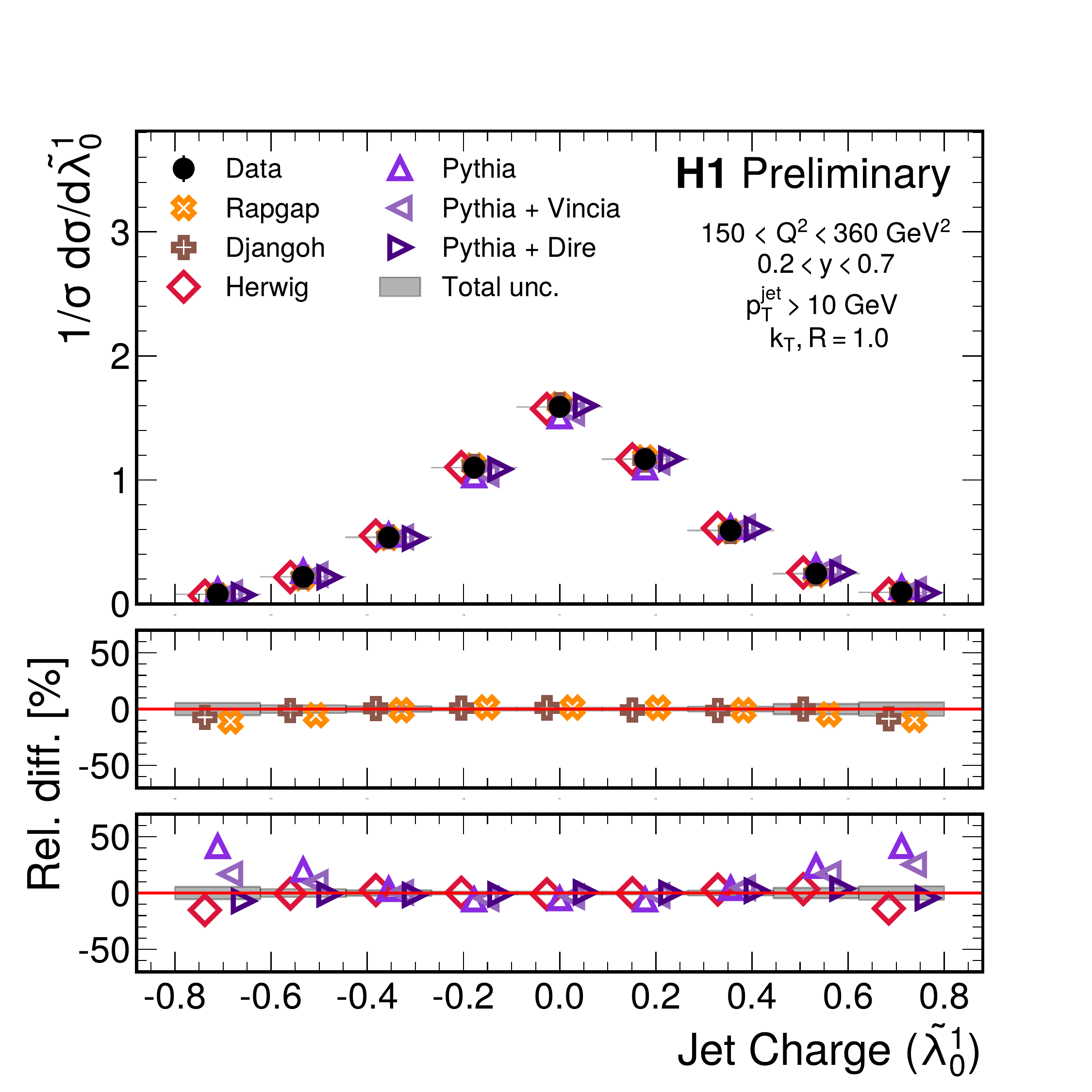}
    \caption{Differential cross section measurements from the H1 Collaboration~\cite{H1:2021wkz,H1prelim-22-031,H1prelim-22-034}, performed in 2021-2022 using data collected in 2006-2007 and an unfolding method proposed in 2019~\cite{Andreassen:2019cjw}.}
    \label{fig:my_label}
\end{figure}

\paragraph{HERA/H1.}  Unlike the $e^+e^-$ experiments at the $Z$ pole (LEP/SLD), the deep inelastic scattering experimental collaborations at HERA continue to exist.  Their software infrastructure has been modernized~\cite{Britzger:2021xcx} and it is still possible to analyze data from two decades ago.  These data are precious and new insights from jet physics and modern machine learning (ML) will yield new and exciting physics results - see e.g. Ref.~\ref{fig:my_label}.  I recently became involved in the H1 experiment because of my interests in quantum chromodynamics (QCD) and in preparation for the upcoming Electron Ion Collider (EIC).  While the H1 data are proprietary, the collaboration welcomes new members.  There is enough institutional knowledge to enable precise analyses, but not enough institutional inertia to discourage new ideas.  There are simulated datasets that are hundreds of times the size of the experimental data and significant institutional knowledge about how to generate more events, even with different generators.  This seems like an ideal model for long term data analysis in HEP.

\paragraph{LEP/SLC.}  In contrast to HERA, the experimental collaborations at LEP and SLD ended a long time ago and their data are not public.  I have been fortunate to have the opportunity to explore the ALEPH dataset from LEP through personal connections (or rather connections of connections) with former collaboration members.  These data are pristine and just like for HERA, there are many new analyses to be performed with modern QCD and ML tools.  However, the software infrastructure has not been modernized and there are no standard tools for generating new events or estimating uncertainties.  With significant effort, some aspects of these tasks can be performed (see e.g. Ref.~\cite{Chen:2021iyj}), as at least some collaboration software is still accessible.  The situation with the SLD collaboration at SLC is even less positive.  As with ALEPH, I have access to the data through personal connections, but there is little or no simulation and the software to produce more simulation disappeared when old computers were discarded.  As there is very little we can do without simulations, this makes it challenging to use these pristine and unique data. 

\section{Resources for Data Preservation}

Everyone agrees that data preservation is critical, however, it is often an unfunded mandate to make this happen.  Resources for data preservation should be available before, during, and after an experiment has ended. Data from any point in the data taking should be analyzable throughout the course of the experiment’s lifetime.  I am impressed by the HERA model for data preservation, which is possible due to a modest investment from DESY.  It seems like national laboratories are well-suited to host (which is more than just store) these important and large datasets and the accompanying software/documentation.  Funding data preservation after an experiment has ended is complicated because project funding is no longer available.  Funding data preservation centers (like the CERN Open Data Portal) at national laboratories seems like a viable path forward.  It may also be possible that a modest investment now could even resurrect the unique datasets taken at LEP and SLC while enough former collaboration members are around, especially as the community prepares for a future $e^+e^-$ collider.

I note that not everything about data preservation is in dire straits.  A number of tools and initiatives are now in place to preserve various data products and analysis logic.  While not “open data” as I have defined it above, these developments are an important success for the future of HEP.  For example, current and future search results in ATLAS are being preserved through the \texttt{RECAST}~\cite{Cranmer:2010hk} system, which makes it straightforward to determine the sensitivity to new models not tested at the time of the original analysis.  For Standard Model measurements, \texttt{Rivet}~\cite{Buckley:2010ar} and \texttt{HepData} are broadly accepted as community standards for preserving analysis logic and final measurement data products.  These have enabled many data re-analyses.

\section{Data Physicists}

It is clear that there will be more and not less open data.  I have tried to argue above that alternatives to open data may be preferable, but I want to now turn my attention to how I believe HEP should react to the growing availability of these data.  In particular, I think the trend in open data provides further and strong evidence for the need for a new type of HEP scientist that is neither an ``experimentalist'' or a ``theorist'' - they are a ``Data Physicist''\footnote{This name was coined by David Shih, who gave an \href{https://indico.fnal.gov/event/22303/contributions/245346/attachments/157349/205798/Snowmass2022_Plenary_Shih.pdf}{inspirational talk} at the recent Snowmass Summer Study.}.  These scientists have the core skills to understand and interrogate data as well as the computational and theoretical background to relate these data to underlying physical properties.  Unlike a traditional ``experimentalist'', a Data Physicist will likely not have extensive (or any) hands-on instrumentation experience and unlike a traditional ``theorist'', they may not have extensive (or any) experience with complex higher-order cross section calculations.  It is not enough to make data easily accessible and well-documented - we also need to train a cohort of scientists who are well-equipped to use these data for science.  In particular, I expect the Data Physicist to have a HEP background, but extensive training in statistics/data science/machine learning and scientific computing.  Software and computing are becoming commensurate with instrumentation and it is important and necessary physics research to work on these topics for maximizing the science from our data.  

How can we create this cohort?  There needs to be career paths (starting in graduate school) for Data Physicists.  This includes degree programs as well as long term funding prospects (that are not all short grants as is typical for computing).  Important institutions like \href{https://www.anl.gov/hep-cce}{HEP-CCE}, \href{https://iris-hep.org}{IRIS-HEP}, the NSF AI Institutes (\href{https://iaifi.org}{IAFAI}, \href{https://a3d3.ai}{A3D3}, etc.) are great examples of interdisciplinary research on software and computing and I strongly support continuing and expanding these initiatives.  However, funding through individual PIs, as is currently the case for other areas of experimental and theoretical HEP, will be important for a sustained effort.

\section{Conclusions and Outlook}

The LHC still has a long and exciting program ahead - only $\mathcal{O}(1\%)$ of the full dataset has been collected.  The future data will open up exciting avenues for research that are not possible with the existing data.  At the same time, the data from the first and second runs of the LHC are unique.  The low instantaneous luminosity of Run 1 may compensate for the improved detectors of Runs 2 and beyond for certain precision measurements.  I would argue that before we make it a priority to ensure that external users can analyze these legacy datasets, we should work hard to make sure that collaboration members are able to analyze them.

It is fantastic that many people are excited about sharing and exploring data.  The reason I became an “experimentalist” in graduate school is precisely because I wanted to interact directly with data.  On the other hand, I have found through the course of my career so far that the labels “experimentalist” and “theorist” can be unnecessarily restrictive.  With a growing need for cross-cutting methodology and the growing availability of open datasets, we need to train and support Data Physicists to make the most of our precious data.  With open minds and the right skill set, we will be ready to make the discoveries of tomorrow.

\section*{Executive Summary:}

\begin{tcolorbox}
\begin{enumerate}
\item Significant short term investment for Runs 1 and 2 and modest long term investment going forward to ensure that all relevant previous data collected by a collaboration are analyzable by the collaboration.  This could be stress-tested with data challenges, including targeted measurements that would benefit from earlier data.  The focus should be on enabling generic data analysis and is complementary to efforts to preserve historical analysis logic and data products. This could benefit from involvement by non-collaboration researchers (see 2 and 3).

\item Make it easier for non collaboration members to engage with the collaboration on data analysis.  This should include ease procedurally as well as financially (see 3).

\item Significant and long term investment (‘base funding’) in Data Physicists at all levels (undergraduate, graduate, postdoc, and tenure-track faculty/staff).  This funding should be different than existing experimental/theoretical sources; data physicists are not `experimentalists' who do not touch detectors and are not `theorists' who do data analysis - they have dedicated training and a focused research program.  Data Physicists are also not engineers, although they are also critical for enabling the science.
\end{enumerate}
\end{tcolorbox}

\section*{Acknowledgements}

I thank Matt Bellis for initiating this white paper (in concert with Ref.~\cite{other}).  Many of the ideas expressed here are the direct result of extensive conversations with Jesse Thaler and Hannah Joo spanning many years.  I have also benefited from enlightening discussions with many other people including Stephen Bailey, Kees Benkendorfer, Krish Desai, Matt LeBlanc, Radha Mastandrea, Jennifer Roloff, and David Shih.  The views expressed here are my own, but I would like to thank the ATLAS Collaboration for years of discussions that have helped form my perspective and for the opportunity to contribute to and benefit from our dataset.  Furthermore, I would like to thank the H1 Collaboration (including Stefan Schmitt and Daniel Britzger) for being so welcoming and Miguel Arratia for connecting me to the collaboration in the first place.  I also thank Tony Johnson and Su Dong for helping me explore the SLD data.  Lastly, I thank my ALEPH data collaborators, in particular, Anthony Badea, Patrick Komiske, Yen-Jie Lee, Eric Metodiev, and  Jesse Thaler.





\clearpage

\bibliographystyle{JHEP}
\bibliography{myreferences}  


\end{document}